Chanhou Lou*

# Judging Data: Critical Discourse and the Rise of Data Intellectual Property Rights in Chinese Courts

**Abstract:** This paper examines how *Sino-judicial activism* shapes *Data Intellectual Property Rights* (DIPR) in China through *Critical Discourse Analysis* (CDA). It identifies two complementary judicial discourses: local courts such as the Zhejiang High People's Court (HCZJ) adopt a *judicial continuation discourse* that pragmatically extends intellectual property norms to data disputes, while the Supreme People's Court (SPC) employs a *judicial linkage discourse* aligning judicial reasoning with national policy and administrative governance. Their interaction forms a *Bidirectional Conceptual Coupling* (BCC) – an "inside-out" projection of local court reasoning and an "outside-in" translation of policy into judicial interpretation by the SPC – that mutually legitimizes and constrains judicial and policy frameworks, balancing the demand for unified market standards with safeguards against platform-driven monopolization. Through cases such as the HCZJ's *Taobao v. Meijing* and the SPC's *Anti-Unfair Competition Interpretation*, the study shows DIPR as an experimental field for doctrinal innovation and institutional coordination in China's evolving digital governance.

# 1 Introduction

On June 22, 2022, the People's Republic of China (PRC) issued the *Opinions on Establishing a Data Base System to Maximize a Better Role of Data Elements*, proposing the establishment of a "data property rights system" (State Council 2022). Scholarly debates on data property rights have primarily focused on legislative experimentation (Li 2017; Yu and Zhao 2019; Liu 2023a; Ma 2024), often neglecting the judiciary's role. Legislative initiatives – such as Article 127 of the *Civil Code* adopted at the National People's Congress (NPC), Article 4 of the *Shenzhen Special Economic Zone Data Regulations*, and Article 12 of the *Shanghai Data Regulations* – remain too abstract to provide a reliable basis for adjudication (National People's Congress 2020; Standing Committee of Shenzhen Municipal People's Congress 2021a, 2021b). Consequently, courts face difficulties in directly applying these provisions to resolve disputes.

---
**\*Corresponding author: Dr. Chanhou Lou**, Macau Fellow Researcher at the University of Macau, 2012 Office, E32, Avenida da Universidade, Taipa, 999078, Macau, E-mail: chlou@um.edu.mo. https://orcid.org/0009-0003-3494-1079

On the other hand, administrative agents highlight registration and regulatory frameworks, including the pilot programs of the China National Intellectual Property Administration (CNIPA) for blockchain-based notarization and public registration (CNIPA 2023; Xu 2025). Yet these initiatives serve primarily symbolic functions, failing to resolve the deeper conceptual debate on whether data should constitute a new intellectual property category (Huang 2024; Sun 2025).

"Data Intellectual Property Rights" (DIPR) as a relatively new term was proposed by Chinese policy discourses, including the *Opinions on the Reform and Innovative Development of Digital Trade*, the *Outline for Building an Intellectual Property Powerhouse (2021–2035)* and the *Notice of Issuing the National Intellectual Property Protection and Use Plan for the 14th Five-Year Plan Period* (State Council 2021a, 2021b, 2024). While it signals an emerging recognition that DPR should be conceptualized within the framework of intellectual property (IP), its precise legal meaning and scope remain vague, leaving room for competing interpretations and institutional experimentation.

In contrast to legislative and administrative dominance of interpretations (Ma 2024; Xu 2025), some Chinese scholarship highlights the judiciary's emerging role (Kong 2024; Yang et al. 2025). Judicial activism naturally emerges amid the societal transformations underway in China (Wang 2006). Zhang (2012) reinterprets the Supreme People's Court (SPC) as a pragmatic institution that, in pursuing its own financial security and sociopolitical standing, acts as a "rational actor" driven by institutional self-interest. The former IP judge at the SPC of the Professor Xiangjun Kong (2015) defines "Sino-judicial activism" as a context-specific form of judicial engagement, combining two dimensions: shaping behavioral and normative standards and enhancing the judiciary's authority in public perception. Earlier, the *Outline of the National Intellectual Property Strategy* emphasized "giving full play to the leading role of judicial protection" and strengthening judicial capacity (State Council 2008), while subsequent opinions in 2018 reaffirmed this leadership role (General Office of State Council 2018a). This perspective in IP adjudication challenges the assumption of "judicial deference" (He 2020). Courts with judicial activism can be equally innovative, actively using equitable reasoning and interpretive techniques to address gaps left by administrative and legislative frameworks.

International debates mirror this tension in IP aspects. Proponents like Lessig (2002) and Schwartz (2004) advocate coupling data protection with intellectual property, while critics such as Reichman and Samuelson (1997) warn against the rigid commodification of data. Some proponents note that emerging UK and historical US case law recognize IP rights in personal data (Trakman et al 2019). Yet most scholarship focuses on the ontological question of whether data qualifies as intellectual property (Cui 2022; Feng 2022; Liu 2023b; Jia 2024; Dong 2025), overlooking the institutional question of who should protect it. This reflects what Hart (1961) calls the distinction between "primary rules" defining rights and obligations and "secondary rules" concerning institutional authority to interpret and adapt them. Aligning with Hart's view, Sino-judicial activism highlights the judiciary's interpretive and standard-setting role in emerging legal fields (Kong 2015, 2024). Yet few studies systematically examine how judicial activism is expressed in DIPR disputes.

To fill these gaps, this study employs Critical Discourse Analysis (CDA) as both a theoretical and methodological tool. Rooted in Foucault's conception of discourse as a regime that produces knowledge and embeds power within institutions such as law (Foucault 1969, 1975), CDA has been widely used to examine how legal discourse constructs social meaning, institutional authority, and power relations (Fairclough 1992). Fairclough extends Foucault's insight by integrating systemic-functional linguistics and critical social theory, developing a three-dimensional CDA model – text, discursive practice, and social practice – that links language to social structures of power (Fairclough 1995).

Fairclough (1995) conceptualizes discourse as a social practice that not only reflects but also shapes institutional and political structures. Similarly, in legal contexts, judicial discourse is not merely descriptive; it performs a constitutive role in defining the scope of legal norms and the boundaries of institutional authority (Goodrich 1987). Scholars such as Bourdieu (1987), Conley and O'Barr (1990), and Mertz (2007) further reveal how legal discourse performs institutional authority, disciplines cognition, and reproduces legitimacy. CDA has been applied to the study of Chinese judiciary (Liao 2009; Han 2011; Wu et al 2022; Sun et al 2025). Ge and Wang (2019) employ CDA to demonstrate that Chinese civil trial discourse embodies interdiscursivity shaped by hybrid institutional ideologies and four decades of judicial reform. Therefore, this makes CDA particularly suited for analyzing DIPR because it reveals how judges frame data disputes as part of broader intellectual property frameworks or, alternatively, as a unique regulatory problem requiring new institutional approaches. Yet, despite its rich potential, no study has applied CDA to examine how Chinese courts construct DIPR, leaving a critical gap that this research seeks to address.

Overall, this paper from CDA examines the subjective dimension of Sino-judicial activism in DIPR: Should the judiciary lead the extension of intellectual property rules to data? And if so, what form does this activism take? The author argues that judicial reasoning around DIPR reflects a deeper structural contradiction: the need for unified standards in the digital economy versus the risk of reinforcing platform monopolies.

This paper is structured into five main parts. Following this introduction, Part 2 outlines the research methodology, highlighting the adoption of CDA and detailing the case selection and sampling strategy. Part 3 delves into the judges' discourses, contrasting the "judicial continuation discourse" employed by local courts with the "judicial linkage discourse" utilized by the SPC. In Part 4, the discussion shifts to the dynamics of the courts themselves, illustrating through cases such as *Taobao v. Meijing* how judicial discourses evolve within different judicial hierarchies. The final part concludes the study, synthesizing insights from CDA to argue that Sino-judicial activism functions through a "Bidirectional Conceptual Coupling" (BCC) that balances internal judicial reasoning with external policy imperatives.

# 2 Materials and Methods

## 2.1 Critical Discourse Analysis as Framework

This study draws primarily on Fairclough's three-dimensional model (text, discourse practice, sociocultural practice) of CDA (Fairclough 1995), which views discourse as simultaneously: (1) Textual – analyzing the language of judgments, reports, and judicial interpretations, including terminology, modality, and rhetorical strategies, including the nuanced use of terms such as *may* or *should* in judicial interpretations (Cheng and Cheng 2014; Li et al 2016; Supreme People's Court, 2022b); (2) Discursive practice (discourse practice) – examining how judicial texts are produced and circulated within the judiciary – such as the SPC, and the Zhejiang High People's Court (HCZJ) – and how they relate to administrative or legislative actors (Supreme People's Court 2021; Zhejiang High People's Court 2022); (3) Social practice (sociocultural practice) – situating judicial discourse within broader governance frameworks, such as the demand for unified standards of a national market with the risk of reinforcing platform monopolies in digital economy (Supreme People's Court 2022c).

## 2.2 Case Selection and Sampling Strategy

he study uses a purposeful case selection strategy to highlight both the local and national judicial discourses shaping DIPR. Two types of judicial texts were analyzed: (1) Local Court Case Law: Local Court Case Law: Zhejiang Province was selected as a pilot jurisdiction due to its robust digital economy and the concentration of platform-related disputes. Cases such as *Taobao v. Meijing*, *Tencent v. Wu Chengrui*, and *Tonghuashun v. Lighthouse Finance* decided by Zhejiang courts provide valuable insights into DIPR (Hangzhou Internet Court 2017; Hangzhou Intermediate People's Court 2018; Zhejiang High People's Court 2019; Zhejiang High People's Court 2018). Comparative and supplementary materials are drawn from cases outside Zhejiang Province, such as *Qihoo v. Baidu*, *Shuju Tang* in Beijing (Beijing High People's Court 2020; Beijing Intellectual Property Court 2024). (2) Supreme People's Court Judicial Interpretations: The judicial interpretation on anti-unfair competition, facial recognition and related SPC publications were analyzed as examples of DIPR, illustrating how the highest court frames data protection within a broader governance narrative (Supreme People's Court 2021; Supreme People's Court 2022b; Supreme People's Court 2022c).

This sampling strategy reflects both the practical accessibility of judicial texts – given that local court decisions are more readily published – and the institutional hierarchy of discourse, where the SPC's interpretive texts function as both judicial and governance signals (Fairclough 1995).

The study also considers secondary materials such as research reports and articles authored by judges in the HCZJ and the SPC, which bridge judicial reasoning and policy interpretation (Zhejiang High People's Court 2022; Qin 2021).

## 2.3 Limitations

While the CDA approach provides valuable insights into the institutional dynamics of judicial discourse, it is primarily qualitative and interpretive. The findings rely on the textual availability of judicial documents, which may be incomplete, especially for privacy-related disputes that are not publicly disclosed (Lou 2023a).

Moreover, this study focuses primarily on Zhejiang as the core research site. This choice is methodologically deliberate: unlike quantitative research, which can compare larger and more dispersed datasets across multiple provinces, CDA is better suited to an in-depth "thick description" of a localized judicial context. Zhejiang provides a uniquely rich sample for such deep analysis, given its status as a leading hub of China's digital economy, its concentration of platform enterprises like Alibaba, and the pioneering judicial practices of courts such as the Hangzhou Internet Court and the HCZJ. These courts have produced more publicly accessible judgments and research reports on DIPR compared to other jurisdictions, making Zhejiang a particularly fertile site for CDA.

Therefore, while Zhejiang offers analytically robust and representative cases for understanding Sino-judicial activism, the findings may not be fully generalizable to other provinces with different economic structures, platform ecosystems, or judicial practices. Although this study also incorporates selected judgments and materials from local courts outside Zhejiang province (Beijing High People's Court 2020; Beijing Intellectual Property Court 2024; Henan High People's Court 2022) Henan High People's Court 2022), these sources primarily serve as supplementary and comparative evidence for the CDA analysis of Zhejiang courts and are therefore insufficient to enable a comprehensive horizontal CDA comparison across different provincial courts in China. Future research could expand this CDA-based approach to other jurisdictions to examine how regional judicial discourses interact with China's broader DIPR governance framework.

# 3 Judges' Discourses

The landscape of Sino-judicial activism is largely shaped by the interpretive contributions of judges at both the local and national levels. Judges from local courts as well as the SPC have engaged in scholarly writing on the emerging concept of data intellectual property, offering a valuable corpus for discourse analysis. This judicial literature enables the identification of distinct interpretive approaches and institutional perspectives.

To capture the layered structure of these discourses, the study examines how judges conceptualize the term DIPR following the hierarchical structure of the judiciary – from local court judges to justices of the SPC. Through comparative analysis of these materials, two general patterns of judicial discourse emerge. Local court judges tend to adopt what may be described as the "judicial continuation discourse," emphasizing continuity with established intellectual property norms. In contrast, the SPC judges are more likely to advance a "judicial linkage discourse," which situates data intellectual property within broader administrative and legal reforms, thereby implying a systemic reconfiguration of legal authority.

## 3.1 Local Court Judges' Straightforward Expression: Judicial Continuation Discourse

At the local level, judges articulate Sino-judicial activism in a pragmatic manner through what can be termed the judicial continuation discourse. A notable example is the HCZJ Joint Research Team's report (2022) titled "On the Intellectual Property Protection of Enterprise Data Rights," published in *People's Judicature*.

Although the report does not explicitly use the term "Sino-judicial activism," it endorses its essence by advocating a sequential path: judicial precedent, followed by administrative regulation and legislative codification. It states, "Overall, we suggest following the path of judicial precedence, administrative follow-up, and legislative summary, formulating data legislation through summarizing judicial experiences and administrative regulation experiences, and enacting specialized laws on data property or data rights." (Zhejiang High People's Court 2022) This marks a shift from judicial restraint toward proactive judicial leadership.

The report further acknowledges that only certain types of data – those meeting thresholds of novelty, creativity, or originality – may qualify for protection under existing intellectual property laws, while raw or unstructured data remain outside the current scope of copyright and patent protection (Zhejiang High People's Court 2022). In this way, the discourse affirms that, even in the absence of legislative or administrative provisions, judicial recognition can extend and adjust existing intellectual property norms to data property – an affirmative expression of Sino-judicial activism. There, the Zhejiang judges identify three core dimensions of judicial subjectivity that inform judges' leading interpretive roles. These include conceptual innovation, normative reasoning, and hierarchical authority in defining emergent legal standards (Zhejiang High People's Court 2022). It means that legal discourse not only reflects existing norms but also actively recontextualizes them to address new social practices (Fairclough 1992).

### 3.1.1 Originality and Intellectual Labor in Compilations

The HCZJ report highlights that many datasets do not meet the originality threshold required for copyright protection. For example, the real-time stock market database of the Shanghai Stock Exchange is described as merely the inevitable outcome of fixed calculation rules, lacking any creative input necessary for protection under copyright law (Zhejiang High People's Court 2022). This exposes a dilemma: as datasets grow more comprehensive, their selection and arrangement become less original, making them harder to protect legally – yet this very comprehensiveness is what gives data its economic value.

Through this reasoning, judicial discourse implicitly challenges the formalist view that protection depends strictly on originality. By recontextualizing the meaning of originality in data property disputes, Zhejiang judges frame intellectual labor in data collection as a form of value deserving recognition. This discursive shift resonates with Brad Sherman and Lionel Bently's historical account of the "shift from creation to object," where nineteenth-century British law transitioned from focusing on creative labor to emphasizing the characteristics of protected works (Sherman and Bently 1999).

By reintegrating "creation" as a normative concept in judicial reasoning, Zhejiang judges extend protection beyond traditional originality, invoking a broader Lockean labor theory of property to recognize effort-intensive data collection as worthy of legal acknowledgment (Locke [1690] 1988). This linguistic and conceptual move exemplifies how judicial discourse shapes emergent norms for data intellectual property, revealing the judiciary's proactive role in filling gaps left by rigid statutory definitions.

### 3.1.2 Unfair Competition and Judicial Discretionary Standards

Zhejiang judges highlight the challenges of applying the overly abstract Article 2 of the Anti-Unfair Competition Law, which risks enabling inconsistent rulings and accusations of "judicial arbitrary discretion" in data disputes (Standing Committee of the National People's Congress 2019). To address this, they develop a structured "five-factor theory" – evaluating data circulation value, incentives for innovation, technological protection, systemic security, and personal information safeguards – as a framework for adjudicating unfair competition in data-related cases (Zhejiang High People's Court 2022).

Through this reasoning, judicial discourse moves beyond statutory formalism by embedding data disputes into broader market governance goals, such as maintaining competitive order and enhancing social welfare. CDA reveals how this discourse links micro-level case judgments with macro-level policy narratives, aligning judicial discretion with national regulatory objectives. It exemplifies how judicial discourse functions as a discursive practice that links local adjudication to broader socio-economic narratives, mediating between textual choices and institutional power (Fairclough 1995).

Judges further refine discretionary standards by distinguishing between non-public, partially public, and fully public data based on accessibility, and between raw and derived data based on processing. Derived data, requiring more intellectual labor, is seen as more deserving of protection. They also reinterpret "commercial morality" under Article 2, explicitly stating that in rapidly evolving sectors, "judges must create and define" industry morality where "wide recognition" does not exist (Zhejiang High People's Court 2022).

These interpretive practices exemplify Sino-judicial activism, showing how courts preempt legislative and administrative gaps by constructing normative benchmarks for unfair competition in data intellectual property.

### 3.1.3 Review of Crawling Protocols and Judicial Objectification

The Robots Exclusion Protocol (crawling protocol) is a unilateral declaration by platforms, functioning as a "no trespassing" sign to limit third-party data crawling. In China's digital economy, this creates conflicts between incumbent platforms – who defend their accumulated data resources – and entrant platforms seeking access.

Zhejiang judges recognize the protocol's utility in reducing coordination costs but caution against its misuse as an anti-competitive tool (Zhejiang High People's Court 2022). Judicial reasoning therefore calls for scenario-based review to assess whether violating such a protocol constitutes unfair conduct, ensuring it is not weaponized to block competition.

The *Qihoo v. Baidu* case illustrates this approach (Beijing High People's Court 2020). Although the HCZJ was not the adjudicating court in this case, its research report evaluated this judgment, offering a unique perspective on how higher people's courts comment on and assess each other's rulings. The HCZJ commented that Baidu's discriminatory application of crawling restrictions violated the protocol's institutional aim – facilitating broader data access for public information sharing – thereby constituting an abuse of regulatory discretion. Through CDA, this judicial discourse reveals how courts reinterpret self-regulatory platform rules in line with public welfare narratives, reframing private data restrictions as a governance issue aligned with national digital policy goals.

Zhejiang judges further challenge the absolute property claims embedded in crawling protocols, shifting from strict liability to fault-based standards. In doing so, the judiciary emphasizes the active circulation of data over passive stockpiling, reasserting that the value of data lies in its dynamic use rather than exclusive ownership.

Therefore, Sino-judicial activism must strike a delicate balance: it should lead the development of the objective theory to achieve national market standardization while maintaining critical distance to avoid judicial discretion being exploited to entrench vested interests. Courts are expected to build predictable, stable rules necessary for a unified digital economy, yet they must remain vigilant to prevent dominant platforms from capturing these rules to cement their monopolistic power. This balance frames judicial discretion as both an enabler of governance and a safeguard against institutional abuse.

Judges' understanding of Sino-judicial activism emerges from two intertwined forces: the spontaneous order of the digital economy and top-down national policy decisions. This hybrid framework of internal and external rules departs from Friedrich Hayek's dichotomy, which privileges internal rules arising from social practices and de-emphasizes externally imposed directives. Hayek argued that internal rules are discovered organically by courts as rules adhered to and are complemented by supplements necessary for their smooth and effective operation (Hayek 1973). However, the Chinese context blends these internalized judicial norms with explicit external policy goals, creating a hybrid model that exceeds the boundaries Hayek envisioned.

The limitations of relying solely on internal rules are evident in the opacity of platform-based economic interactions. Crawling protocols, for example, are highly technical and invisible to the public, while even professionals face algorithmic "black boxes" and fragmented programming tasks that obscure their systemic functioning. This complexity eliminates the universality and abstraction Hayek assumed for internal rules in a spontaneous legal order.

Moreover, platform "private power" deliberately creates external rules – organizational protocols that mimic customs but are strategically designed to favor incumbents. Kirzner's "Optimism–Pessimism Knowledge Dichotomy" explains this gap: users often lack sufficient knowledge to contest platform practices or even perceive judicial bias, leading to a pessimistic disengagement. This "Beta knowledge problem" makes it impossible for courts, operating solely under spontaneous internal rules, to perceive users' unjust reactions and recognize whether courts' decisions are being shaped by platform influence (Kirzner 1992).

For this reason, the judiciary requires purposeful institutional guidance – organizational rules and national directives – to consciously identify and mitigate risks of capture. This dual dependence shows that Sino-judicial activism is not merely a passive discovery of social norms, but an active construction of discretionary standards aligned with broader governance goals.

In sum, judicial continuation in Sino-judicial activism draws from internal judicial reasoning while inevitably invoking external interventions. What appears as straightforward "judicial continuation discourse" by local courts becomes an "inside-out" trajectory of governance, expanding from courtroom reasoning into the broader institutional separation of judiciary and business. Sino-judicial activism thereby balances market standardization with safeguards against the monopolization of legal norms by private power.

## 3.2 Supreme Court Judges' Metaphor: Judicial Linkage Discourse

In contrast to the straightforward "judicial continuation discourse" previously discussed, how do SPC judges conceptualize Sino-judicial activism? A closer reading suggests that their articulation takes a more metaphorical form, aligning with what the author terms the "judicial linkage discourse." For instance, on April 29, 2021, *People's Court Daily* published a theoretical article by Yuanming Qin, Senior Judge of the Intellectual Property Tribunal of the SPC, entitled "Research on Legal Issues Related to the Judicial Protection of Data Property Rights and Intellectual Property Rights" (Qin 2021).

This metaphorical register reflects Fairclough's observation that discourse at the apex of institutional hierarchies tends to naturalize policy priorities as taken-for-granted truths (Fairclough 1995). Here, SPC discourse shifts from explicit normative reasoning to metaphorical alignment, illustrating a social practice in which judicial language reinforces national governance imperatives.

Judge Qin's article adheres to the ontological-to-subjective argumentative structure previously analyzed but renders the subjectivity of Sino-judicial activism metaphorically rather than explicitly. First, by framing the issue within the scope of "judicial protection," the title itself symbolically shifts the discourse from legislative experimentation to judicial authority. Second, the article argues from an ontological position that "data property rights have strong intellectual property attributes," thereby affirming the applicability of intellectual property norms to data property rights. This mode of reasoning directly resonates with the SPC's prior interpretation of "the law of intellectual property protection" as the jurisprudential basis for Sino-judicial activism. Thus, the reasoning adopted in this article metaphorically yet unequivocally advances the view that Sino-judicial activism is in a phase of "emergence."

### 3.2.1 Judicial Linkage with Administrative Explanation

Although the judge-author does not explicitly use the term *Sino-judicial activism*, the concept is implied toward the end: *"The judicial protection of data property rights and intellectual property rights should also be connected with administrative protection to quickly form a comprehensive protection pattern for data property rights"* (Qin 2021). This reflects a structural approach echoed in national policy, such as the *O. for Building an Intellectual Property Powerhouse (2021–2035)* – hereafter "Outline (2021–2035)" – which calls for clarifying administrative–judicial responsibilities, improving linkage mechanisms, and forming a "protection synergy" (State Council 2021a). Similarly, the *National Intellectual Property Protection and Use P. for the 14th Five-Year Plan Period* – hereafter "14th Five-Year Plan" – emphasizes unifying administrative enforcement and judicial adjudication standards, creating an organically linked framework (State Council 2021b).

Textually, the "linkage discourse" appears to promote judicial restraint – engaging with administrative protection in a coordinated but limited way. Yet in practice, intellectual property adjudication suggests a reversal of this restraint. For example, in trademark invalidation disputes, courts often override administrative findings and directly review new grounds beyond those raised in administrative proceedings (Lou 2023b), as seen in the *"Da Yima"* trademark case where the SPC revoked the mark under Article 10 of the Trademark Law. In another word, courts sometimes actively override or go beyond administrative findings in IP cases, rather than deferring to administrative decisions.

Thus, the linkage discourse is judicial-led rather than administrative-led. Policy documents support this view: the *China Program for Judicial Protection of Intellectual Property Rights (2016–2020)* calls for "fully utilizing the institutional and systemic advantages of judicial protection" while supervising and regulating administrative actions (Supreme People's Court 2017). Likewise, the *Opinions on Several Issues concerning Heightening Reform and Innovation in the Field of Intellectual Property Adjudication* states both the need to "play the leading role of judicial protection" and to "promote the unification of administrative and judicial standards" (General Office of State Council 2018a).

This reasoning also responds to the criticism of judicial reliance on general clauses in emerging data disputes. Qin highlights that the rapid growth of the digital economy forces courts to apply broad provisions, such as those in the *Anti-Unfair Competition Law*, even without clear legislative bases (Qin 2021). To address this, he recommends judicial interpretation to clarify and unify standards, rather than waiting for new statutory laws. By prioritizing judicial interpretation over legislative action, the article implicitly advances Sino-judicial activism, showing the judiciary as the primary actor in defining data property protections.

### 3.2.2 Confirmation of the Sino-Judicial Activism Metaphor

Does the metaphor of Sino-judicial activism find confirmation? It appears so, though now expressed more implicitly through national policy and judicial interpretation. A linguistic comparison of national decisions reveals a shift. For example, the *Outline of the National Intellectual Property Strategy (2008)* prominently used the phrase "play the leading role of judicial protection of intellectual property," but this phrase no longer appears in Outline (2021–2035 or the 14th Five-Year Plan (State Council 2008; State Council 2021a, 2021b). Similarly, terms like "data intellectual property" and "Sino-judicial activism" no longer synchronously appear in a single authoritative text in 2021, suggesting a move toward a more metaphorical articulation of judicial leadership.

However, this does not imply that Sino-judicial activism has been abandoned. Instead, the terminology has shifted to a metaphorical register while remaining active in judicial policy and practice. For instance, on September 12, 2019, *People's Court Daily* published an article by Hu Gongyi, then Acting President of Yiwu Primary People's Court in Zhejiang province, urging courts to "fully play the leading role of judicial protection of intellectual property" (Hu 2019). Likewise, a 2022 white paper by the Henan High People's Court reiterated that courts across the province fully play the leading role of judicial protection of intellectual property, implementing strict intellectual property protection (Henan High People's Court 2022). These statements, especially from local courts and judicial leaders, demonstrate that Sino-judicial activism continues as a guiding principle, albeit metaphorically expressed rather than explicitly stated in national-level texts.

This metaphorical presence is further reflected in judicial interpretation. On March 17, 2022, an SPC representative explained that the Court is "exploring the improvement of data intellectual property rights protection rules" in network-related unfair competition disputes (Supreme People's Court 2022a). This interpretation incorporates "business ethics" under Article 2 of the *Anti-Unfair Competition Law* and explicitly articulates it in Article 3 of the judicial interpretation (Supreme People's Court 2022b).

Article 3, paragraph 1 implies a "business subject theory," defining business ethics as norms "universally followed and recognized in specific business fields." Paragraph 2 introduces a "stakeholder theory," extending considerations beyond businesses to consumers, market order, and public interest. Paragraph 3 again reflects the business subject theory, referencing "professional codes, technical standards, and self-regulation agreements" by industry associations and self-regulatory bodies (Supreme People's Court 2022b).

This interplay of theories generates a metaphorical tension. Paragraphs 1 and 3 use "may," indicating discretionary judicial flexibility aligned with Sino-judicial activism, while paragraph 2 uses "should," binding discretion more firmly to stakeholder interests and broader market objectives. Thus, even when judicial interpretations appear to limit discretion textually, they indirectly reinforce Sino-judicial activism by highlighting the judiciary's interpretive authority over emergent data disputes.

To better understand how judicial language shapes normative expectations, insights from modality studies in legal discourse are particularly relevant. Cheng and Cheng (2014) demonstrate how epistemic modality in court judgments reflects varying degrees of judicial certainty and discretion, and Li et al (2016) show that "should" carries stronger deontic force in legislative discourse, aligning interpretation more firmly with policy goals.

This interplay of theories creates a layered modality. Paragraphs 1 and 3 employ "may," signaling a weaker, more discretionary deontic force that aligns with Sino-judicial activism, while paragraph 2 uses "should," imposing a stronger normative expectation tied to broader stakeholder interests. Applying these insights, the modal shift from "may" to "should" in Article 3 reflects a dynamic balance: while judicial interpretations appear to constrain discretion textually, they implicitly reinforce Sino-judicial activism by preserving the judiciary's interpretive authority over emergent data disputes. This demonstrates how modality not only encodes obligation but also negotiates the power-knowledge relationship between judicial flexibility and normative alignment.

In conclusion, the absence of explicit terminology in national policies does not mark a retreat but a shift toward metaphorical judicial leadership. Sino-judicial activism persists as a normative force, primarily revealed through local courts, judicial interpretation, and the flexible application of business ethics in emerging data governance disputes.

### 3.2.3 Diversity in the Objective Theory of Sino-Judicial Activism

Why is a metaphorical discourse used in Sino-judicial activism? This stems from tensions within the objective theory itself. As discussed earlier, national policy (external rules) intervenes in judicial practice to prevent the objective theory from being captured by platforms – a risk SPC judges clearly recognize given recent regulatory shifts.

For example, the *Guiding Opinions on Promoting the Regulated and Sound Development of the Platform Economy* (General Office of State Council 2019) called for both "inclusive" and "prudent" regulation. Later, the Central Political Bureau emphasized the need to "form a strong domestic market … and strengthen anti-monopoly and prevent the disorderly expansion of capital" (People's Daily 2020a). Similarly, the Central Economic Work Conference (People's Daily 2020b) supported innovative development for platforms but simultaneously called for stricter legal norms, especially on monopoly identification, data collection and use, and consumer rights protection.

As further evidence, the *Reform Plan on the Division of Central and Local Fiscal Powers and Expenditure Responsibilities in the Field of Intellectual Property Rights*, issued by the General Office of the State Council in 2023, explicitly assigns the formulation of DIPR protection rules and the promotion of industry standards for DIPR to the central fiscal authority, with related expenditures borne by the central government (General Office of State Council 2023). This implicitly indicates how the discourse of building a unified domestic market shapes the governance of DIPR.

From the judiciary's perspective, these directives reveal the inherent contradiction of the objective theory: the domestic unified market offers functional benefits but risks external capture by platform capital. Consequently, national regulation on platforms has pushed the terminology of Sino-judicial activism into a more metaphorical register.

This is evident in the judicial interpretation of "business ethics." If based solely on the *business subject theory* – which recognizes only platform practices – it risks reinforcing platform dominance. To counter this, the interpretation also incorporates the *stakeholder theory,* which extends discretionary consideration to consumers, market order, and public interest. Embedding both theories within the same clause metaphorically balances the judiciary's interpretive flexibility with the state's objective standard (Supreme People's Court 2022b).

It is not clear whether this metaphor is intentional or an unconscious judicial articulation. Yet its discovery is significant. Notably, the article by the SPC judge

Qin published on April 29, 2021 – before the national policy documents, Outline (2021–2035) explicitly referred to "data intellectual property" published on September 22, 2021 – demonstrates a preemptive judicial induction of internal rules from the digital economy rather than mere reactive alignment with policy (Qin 2021; State Council 2021a).

This shows that Sino-judicial activism emerges through a dual process: shaped by top-down national decisions while also induced bottom-up through judicial adjudication of the digital economy's spontaneous order. In this way, SPC judges metaphorically construct a *judicial linkage discourse*, where external organizational rules between judiciary and administration lead inward toward internal judicial rules – a discursive "outside-in" trajectory.

# 4 Courts' Dynamics

As discussed earlier, the transition from ontology to subjectivity in Sino-judicial activism is inseparable from the question of recognition – namely, from "What is recognized?" to "Who performs the recognition?" In doctrinal terms, this raises a jurisdictional issue: Which courts are authorized to adjudicate disputes involving DIPR? Furthermore, do these courts' judges articulate a *judicial linkage discourse* or a *judicial continuation discourse*? Mapping these "courtroom dynamics" reveals how different levels of courts shape the landscape of Sino-judicial activism from CDA, explaining why SPC and local court judges display distinct conceptual preferences.

To illustrate this, the Zhejiang judges' research report is particularly instructive, as it publicly discloses the members of the Joint Research Group of the HCZJ. This disclosure suggests that these judges have practical adjudicatory experience with recognition rules involving data intellectual property rights and that their courts hold relevant jurisdiction. Without such involvement, it is unlikely these courts would participate in such a specialized research project (Zhejiang High People's Court 2022).

An examination of the 12 identified members reveals two key patterns. First, in terms of court level: four members belong to the Zhejiang Provincial High Court; four come from intermediate courts in Hangzhou, Ningbo, Huzhou, and Shaoxing; three are from primary courts in Yuhang District (Hangzhou), Deqing County (Huzhou), and Jiaojiang District (Taizhou); and one member represents the Hangzhou Internet Court, a specialized court under basic court jurisdiction. Second, in terms of professional specialization: six members are from the intellectual property adjudication field (including judges and assistants), and one member is linked to internet adjudication.

These patterns demonstrate that Sino-judicial activism is shaped through a continuation of legal interpretation across different court levels. By tracing these judicial interactions, one can see how recognition of data intellectual property rights develops incrementally from local courts upward, reinforcing the layered dynamics between *judicial continuation* at the grassroots level and *judicial linkage* at the SPC level.

## 4.1 Reflections on Judicial Continuation in the Taobao v. Meijing Case

Qiong He, the deputy head of the Intellectual Property Tribunal of the Zhejiang Provincial High People's Court and a member of the research group, presided over the 2019 retrial of the *Unfair Competition Dispute between Anhui Meijing Information Technology Co., Ltd. and Taobao (China) Software Co., Ltd.* – hereafter referred to as the "*Taobao v. Meijing* Case." This case was not only cited as a data intellectual property rights case in the Zhejiang research report but was also recognized by the SPC as an "Outstanding Case of the National Court System in 2019." Given its representative nature, the case provides a valuable starting point for reflecting on the "judicial continuation discourse" within the context of local court dynamics (Hangzhou Internet Court 2017; Hangzhou Intermediate People's Court 2018; Zhejiang High People's Court 2019).

### 4.1.1 The Conveyor Belt of Judicial Continuation

The case *Taobao v. Meijing* offers a clear illustration of judicial continuation in data-related disputes. Taobao, an Alibaba-owned platform, alleged that Meijing encouraged users to share sub-accounts of Taobao's *Business Consultant* service through the "Gugu Mutual Assistance Platform" and "Gugu Business Consultant Crowdfunding" sites, unlawfully extracting data and constituting unfair competition. The dispute progressed through two trials and a retrial application.

First, the Hangzhou Internet Court ruled for Taobao, adopting the "original data–derived data dichotomy" and judicially extending it into a tripartite framework: "network user information–original network data–derived data products." It recognized Taobao's *competitive property interests* in the derived data but, citing legislative gaps, declined to "confirm" full property rights. This nuanced language signaled judicial restraint while implicitly affirming the proprietary nature of derived data (Hangzhou Internet Court 2017).

Second, the Hangzhou Intermediate People's Court upheld this reasoning, emphasizing that derived data – analytical outputs like trends and rankings – were no longer directly tied to user information and thus free from "personality" entanglements. This reasoning reduced transaction costs and clarified property interests, aligning with broader intellectual property doctrine on balancing proprietary and personality rights (Hangzhou Intermediate People's Court 2018).

Third, the HCZJ dismissed Meijing's retrial request but further advanced the framework by invoking *creative labor theory*, assessing whether Meijing contributed original labor to justify its use. Drawing on Coase's "reciprocal nature of the problem," the court reframed the issue as whether the law should permit Meijing to harm Taobao or vice versa, ultimately finding no creative contribution from Meijing and thus no justification for its conduct (Coase 1960; Zhejiang High People's Court 2019).

Together, these courts formed a "conveyor belt" of judicial continuation: the Internet Court initiated the tripartite principle, the Intermediate Court refined it by disentangling personality claims, and the High Court enriched it through creative labor theory, embedding fairness-based balancing within unfair competition adjudication. Fairclough (1992) notes that discursive chains – where texts are sequentially interpreted and reinterpreted by different actors – create cumulative authority. The 'conveyor belt' of judicial continuation in *Taobao v. Meijing* exemplifies such a discursive chain, where lower-court reasoning is recontextualized upward, producing an evolving doctrinal narrative.

### 4.1.2 The Universality of Objective Theory

Under the "conveyor belt" model, judicial continuation does not inherently privilege dominant platforms like Taobao, even though it originates from their operational practices. Instead, Sino-judicial activism abstracts platform self-regulation into universal adjudicatory standards applicable across platforms. This addresses concerns of a "platform-captured judiciary" within the judicial continuation discourse.

Coase's (1960) "reciprocal nature of the problem" explains this neutrality: if a subsequent entrant like Meijing engages in creative labor, it could equally qualify for protection under the tripartite framework. Thus, the continuation of the tripartite division does not necessarily strengthen Taobao's unfair competition claims but may validate Meijing's counterclaims of *legitimate harm*. This reciprocity mitigates risks of favoritism by recognizing *mutual legitimacy* between incumbents and challengers, generating a normatively universal rule.

Moreover, inter-platform litigation extends Sino-judicial activism's universal applicability to platform–user disputes. While ordinary users have limited participation in self-regulation, litigation between platforms indirectly safeguards their interests by compelling compliance. As Hayek (1973) notes, competition itself acts as a corrective oversight mechanism.

This dynamic is evident in *Taobao v. Meijing*. Taobao needed to legitimize its user data collection as a foundation for its derived data products, avoiding the "fruit of the poisonous tree" dilemma, where unlawfully sourced data could invalidate subsequent property claims. This focus on personal information protection not only settled the dispute but also educated non-parties (users) on the legality of data practices, incentivizing better compliance in future cases.

Thus, *Taobao v. Meijing* represents a reflexive turn in the judicial objective theory, highlighting that while judicial continuation carries risks of platform capture, such risks are moderated through multi-level court interaction. The "conveyor belt" model shows that Sino-judicial activism can evolve toward institutional continuity that remains normatively desirable, supporting an optimistic view of its role in shaping data intellectual property rights.

### 4.1.3 The Hierarchy of Power-Knowledge

This optimistic view of Sino-judicial activism highlights a critical inference: court-led judicial continuation carries "continuation power." It can extend beyond platform self-regulation to statutory provisions without being constrained. Generalizing from the *Taobao v. Meijing* "conveyor belt" model, this continuation power strengthens up the judicial hierarchy – from the Hangzhou Internet Court to the Zhejiang Provincial High Court, where reflexive reasoning such as the "creative labor theory" further advanced the doctrinal framework (Zhejiang High People's Court 2019).

This layered progression shows that local courts generate a balanced, cumulative discourse of judicial continuation. Yet if the focus narrows to a single lower court, the continuation discourse risks "collapsing," as limited judicial authority or knowledge may fail to sustain doctrinal innovation. This dynamic reflects Michel Foucault's insight: power and knowledge directly imply one another; there is no power relation without the correlative constitution of a field of knowledge, nor any knowledge that does not presuppose and constitute power relations (Foucault [1975] 1979). The author terms this interaction the "hierarchy of power–knowledge."

While Foucault theorized power and knowledge as diffuse, Fairclough's model operationalizes it institutionally: textual innovations are constrained by discursive practices tied to hierarchical authority, and these reflect the social practice of embedding judicial power within state structures (Fairclough 1995). Thus, the vertical court hierarchy shapes how continuation discourse can evolve without collapsing, illustrating how judicial language both produces and is constrained by institutional arrangements.

Two mechanisms explain this hierarchy. First is the vertical responsibility relationship in China's judicial system: the lower a court, the more its decisions – especially innovative continuations – are vulnerable to review or reversal by higher courts. This institutional risk curtails doctrinal experimentation. For example, the Hangzhou Internet Court's rulings are reviewable by both the Hangzhou Intermediate Court and the Zhejiang High Court. By contrast, the High Court's retrial decisions face far less scrutiny, enabling broader continuation power.

Second is horizontal knowledge capacity – the specialized epistemic authority of courts with intellectual property judges. While specialization can enhance continuation power, in data-related IP disputes this factor is largely neutralized because each tier of court already maintains IP tribunals. Hence, horizontal expertise exerts minimal influence compared to vertical responsibility.

In summary, vertical hierarchy primarily shapes judicial continuation power. This helps counter fears of a "platform-captured judiciary," as higher courts – less exposed to local platform influence – can refine and correct lower court reasoning. Yet this reasoning encounters limits when comparing provincial high courts and the SPC. While the SPC should theoretically possess the strongest continuation power, empirical evidence shows SPC judges adopt a more cautious approach. Rather than extending judicial reasoning outward (*internal-to-external*), they tend toward an *external-to-internal* judicial linkage discourse, harmonizing judgments with national and administrative policy.

This divergence suggests the need to reposition the SPC within the broader Sino-judicial activism landscape, highlighting its unique mediating role between local judicial experimentation and top-level policy integration.

## 4.2 The Judicial Linkage Characteristics of the Supreme People's Court

Why do judges of the SPC favor a "judicial linkage discourse" rather than a "judicial continuation discourse"? One explanation lies in the organizational structure of appellate jurisdiction. What the author terms the "exhaustion of appellate jurisdiction" reflects the procedural reality that data intellectual property disputes rarely ascend to the SPC. This scarcity limits the Court's capacity to engage in case-by-case doctrinal refinement typical of lower-level courts.

As a result, the SPC adopts a discourse focused on harmonizing judicial reasoning with administrative and legislative frameworks, rather than continuing doctrinal reasoning through successive cases. In contrast to provincial high courts, which build incremental jurisprudence through individual rulings, the apex court emphasizes discursive alignment with policy objectives, positioning itself as a stabilizing actor within the broader field of Sino-judicial activism.

### 4.2.1 The Exhaustion of Appellate Jurisdiction in DIPR

It is important to note that the Chinese court system follows a two-tier finality rule, meaning that a judgment rendered by the appellate court (second instance) becomes final and legally effective. Whether a case subject to a second-instance effective judgment will be reopened through a retrial is at the discretion of the higher court, thereby constituting a privilege granted by the higher court rather than a right of the parties.

The SPC adopts a differentiated approach to intellectual property adjudication, structuring its jurisdiction through distinct procedural "tracks" that categorize cases by their technical complexity and level of appeal. This structure determines whether a case can exhaust its appellate avenues and reach the SPC. For example, "technical intellectual property cases" – such as those involving patents or monopolies – are typically assigned to intermediate people's courts and intellectual property courts, which apply specialized expertise.

The 2020 amendment to the *Provisions on the Jurisdiction of Cases of the Beijing, Shanghai, and Guangzhou Intellectual Property Courts* explicitly placed first-instance civil and administrative disputes concerning patents, monopolies, and similar technical matters under the jurisdiction of these courts (Supreme People's Court 2020). Moreover, certain technical disputes may be brought directly before the IP Court of the SPC – a newly established sub-institution under the SPC – through leapfrog appeals (SPC 2023).

However, DIPR cases rarely qualify as "technical intellectual property cases." Instead, they are usually classified under causes such as unfair competition or copyright infringement, which remain within the purview of basic courts like the Hangzhou Internet Court or Yuhang District People's Court. For instance, among the cases tried by the research group members-judges of the HCZJ, *Taobao v. Meijing* and *Tencent v. Wu Chengrui* were treated as "pure unfair competition disputes" and thus confined to the lower court tracks, namely the primary courts (Hangzhou Internet Court 2017; Primary Court of Yuhang District Hangzhou 2020). Exceptional cases – such as *Tonghuashun v. Lighthouse Finance* – were elevated to intermediate courts as the first-instance courts only due to high monetary thresholds or the "precisely elevating significant cases" principle under the reform measures (Zhejiang High People's Court 2018).

As a result, most DIPR disputes remain trapped at the basic-court level, with limited opportunities for appellate exhaustion. This institutional design significantly restricts the SPC's ability to develop judicial continuation discourse through individual rulings in this area. Consequently, SPC judges tend to favor the "judicial linkage discourse," aligning their reasoning with administrative and legislative policy rather than engaging in incremental doctrinal development as seen in provincial high courts. It represents a shift from textual practice (case-specific reasoning) to a more monological discursive practice that signals policy alignment, reflecting the social practice of reproducing governance priorities (Fairclough 1995).

### 4.2.2 The External Linkage Function of Judicial Interpretation

The judicial linkage discourse finds one of its clearest expressions in the SPC's practice of judicial interpretation, as exemplified by the *Judicial Interpretation on the Anti-Unfair Competition Law* (Supreme People's Court 2022b). A key question is how such interpretations are understood by SPC judges as mechanisms of external linkage – that is, as a means of integrating judicial authority with broader administrative and legislative frameworks. Crucially, no judicial interpretation explicitly defines data intellectual property or provides binding guidance on how to extend intellectual property norms to issues of data ownership.

Yet, this absence does not mean that judicial interpretations lack metaphorical implications. For instance, doctrines such as the *business subject theory* and the *stakeholder theory* embedded in the *Judicial Interpretation on the Anti-Unfair Competition Law* reflect the tensions and supports of Sino-judicial activism (Supreme People's Court 2022b). Moreover, judicial interpretations indirectly guide data intellectual property disputes. In the *Taobao v. Meijing* case, the court first assessed the legality of personal data collection under privacy protection rules before addressing claims of unfair competition (Hangzhou Internet Court 2017). This illustrates how personal information protection operates as a preliminary standard: only if personal data is lawfully collected can it form the basis for further claims of data rights.

This logic is particularly evident in the *Judicial Interpretation on Facial Recognition*, issued by the SPC on July 28, 2021, just weeks before the promulgation of the *Personal Information Protection Law* on August 20, 2021 (Supreme People's Court 2021). Notably, Article 5 of the interpretation introduced five exemptions for the reasonable use of personal information – far exceeding the two scenarios listed in Article 1,036 of the *Civil Code* yet differing from the six exemptions later codified in Article 13 of the *Personal Information Protection Law* (National People's Congress 2020; Standing Committee of NPC 2021).

Why did the SPC act preemptively rather than wait for the new legislation? Two explanations exist. First, it is unlikely the SPC was unaware of the legislative draft, as Article 13 had been included since October 2020 and was circulated through formal consultations. Second – and more persuasively – the SPC acted from a stance of judicial activism: anticipating that determining the reasonable use of personal information is a prerequisite for establishing platform data property rights. By issuing an interpretation ahead of the law, the SPC provided a creative judgment that bridged the Civil Code's personal information rules with future data ownership claims – metaphorically linking privacy law to data intellectual property (Lou 2020).

However, the *Judicial Interpretation on Facial Recognition* has seen "cold application" in practice. Based on the author's research, only one publicly available case cites it – a local dispute over a property management company forcing homeowners to use facial recognition systems without consent. Media reports on its application are also sparse, suggesting most disputes related to facial recognition are resolved administratively rather than judicially.

This reveals that the interpretation's primary function is external linkage, not internal adjudicatory use. It serves as reference material for administrative regulators and corporate compliance, rather than as an actively applied judicial norm. Many companies list the interpretation as part of their compliance policies, demonstrating its external signaling effect. Moreover, under the "civil–administrative–criminal three-in-one" reform of quasi-intellectual property cases, such interpretations may indirectly influence administrative decisions as the decisions may be reviewed by administrative litigations (Supreme People's Court 2017).

In sum, the *Judicial Interpretation on Facial Recognition* metaphorically reflects Sino-judicial activism by extending judicial influence beyond courts – aligning privacy protection with data intellectual property debates and embedding judicial reasoning into the broader regulatory architecture.

### 4.2.3 The Limitations of Administrative Protection and the Judicial Equity

The judicial interpretation discussed earlier primarily serves an external linkage function, fostering institutional communication and normative persuasion. This raises a central question: why is judicial activism, rather than administrative regulation, considered the more desirable approach for safeguarding data intellectual property and advancing a unified digital economy? This question is especially relevant for the SPC as it addresses the concerns of administrative agencies. Two principal considerations emerge.

First, administrative agencies – most notably the CNIPA – have demonstrated sustained interest in data intellectual property. At a State Council policy briefing on November 1, 2021, CNIPA Commissioner Shen Changyu explicitly referenced the 14th Five-Year Plan (State Council 2021b), affirming the goal to accelerate the development of data intellectual property protection rules. Given this administrative engagement, a key question arises: why should the post hoc regulation of DIPR be addressed through judicial activism rather than relying solely on administrative procedures, which already cover preliminary issues like personal information protection?

Second, the administrative approach itself remains limited in scope. For example, at a 2022 press conference, CNIPA announced pilot projects for data intellectual property protection in Zhejiang, Shanghai, and Shenzhen (CNIPA 2023). After that, the *Measures (Trial Implementation) for the Registration of Data Intellectual Property Rights of Zhejiang Province* was issued (Zhejiang Provincial Administration for Market Regulation 2023). In 2024, Zhejiang Intellectual Property Research and Service Center established a blockchain-based notarization platform in Zhejiang, designed with technical support to issue notarization certificates for DIPR assets (Xinhua News Agency 2024). However, such "notarization registration" merely provides publicity like copyright pre-registration and lacks the legal certainty associated with patent registration (Sun 2025), as illustrated by the *Suju Tang* case (Beijing Intellectual Property Court 2024). If accurate, this indicates that the administrative approach focuses on registration but struggles to provide substantive adjudicatory solutions in defining the essence of DIPR.

On one hand, administrative bodies also face a legislation–enforcement paradox. Creating new statutory law for DIPR would require fundamental reforms in property, civil, and economic law, which under Article 11 of the *Legislation Law* must be enacted by the NPC and its Standing Committee (National People's Congress 2023). Without explicit legislative authorization, the State Council – which oversees and directs the CNIPA – cannot legislate independently (Art. 12). Conversely, attempting to stretch existing intellectual property laws to cover data would risk legality challenges under Article 70 of the *Administrative Litigation Law* for "misapplication of laws," "abuse of power," or "obvious inappropriateness"(Standing Committee of NPC 2017).

Moreover, while Article 2 of the *Anti-Unfair Competition Law* provides a general principle relevant to data disputes, Article 17 (2) clarifies that it is designed for judicial – not administrative – application (Standing Committee of the National People's Congress 2019). The recent restructuring of State Council agencies further complicates jurisdiction: copyright-related DIPR disputes such as originality falls under the National Copyright Administration (General Office of State Council 2008), while platform-related DIPR disputes such as unfair competition fall under the State Administration for Market Regulation, which supervises CNIPA (General Office of State Council 2018b). Thus, the administrative agents are different.

On the other hand, registration mechanisms in the administrative vision are inherently premature. As Sherman and Bently (1999) argue in tracing the evolution of intellectual property from "creation to object," registration merely suspends debates over the creative essence of intangible property by focusing on object description. Even in copyright law, where registration is tied to the object, disputes over intellectual labor persist and are often "captured by new object types." In the case of data, its conceptual "creative essence" remains unresolved. As Zhejiang judges have noted, the originality of data compilations is still evolving (Zhejiang High People's Court 2022). Thus, introducing a registration system at this stage would be like the Chinese idioms – "carving a boat to find a sword" – misaligned with the fluid, transformative nature of data.

In summary, the contradictions within the administrative framework highlight the desirability of judicial activism. Judicial techniques of equity are more flexible and context-sensitive, enabling courts to extend intellectual property rules to data through nuanced recognition and adjustment. Judicial activism in DIPR manifests through imaginative analogies and innovative reasoning, filling the gaps left by rigid administrative processes. In this way, the judicial–administrative linkage becomes not merely complementary but also a reflection of the evolving conceptual framework of Sino-judicial activism.

# 5 Conclusions

This study demonstrates that Sino-judicial activism is central to the evolving governance of DIPR in China. From CDA, it reveals how judicial discourse both reflects and constructs institutional authority, bridging gaps where legislation and administration remain incomplete. CDA uncovers two complementary discourses: the "judicial continuation discourse" of local courts, which pragmatically extends intellectual property norms to data products through case adjudication, and the "judicial linkage discourse" of the SPC, which aligns judicial reasoning with national policy and administrative regulation. Together, these discourses create a "bidirectional conceptual coupling" (BCC) theory between case-based internal reasoning and externally oriented policy coordination.

BCC, as illustrated in Figure 1, is the dynamic interaction between inside-out judicial discourse and outside-in policy discourse, whereby they mutually legitimize and constrain each other, balancing the demand for unified market standards with safeguards against platform-driven monopolization. Therefore, the judicial continuation discourse is stronger on text and discursive practice, while the judicial linkage discourse is more about social practice. Together they create the BCC dynamic, and Fairclough's CDA three dimensions reasonably explains this multi-level interaction (Fairclough 1995).

The *Taobao v. Meijing* case illustrates how local courts develop objective standards for DIPR through layered adjudication, balancing data protection with competition and avoiding platform monopolies. By contrast, the SPC, limited by appellate jurisdiction, engages DIPR indirectly, issuing metaphorical judicial interpretations – such as the Facial Recognition Interpretation – that link personal information protection to data property disputes while providing guidance for administrative regulators.

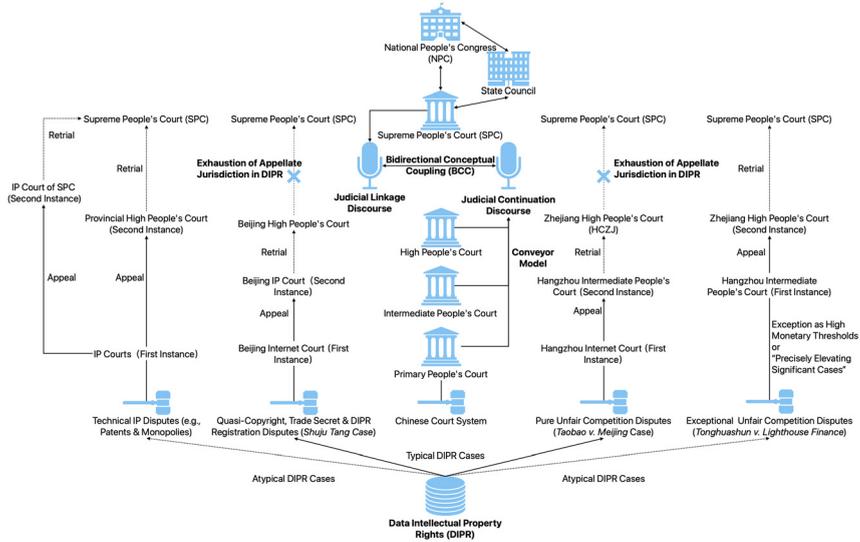

**Figure 1:** Bidirectional conceptual coupling (BCC) of data intellectual property rights (DIPR) in Chinese courts, based on critical discourse analysis (CDA).

Even where expressed cautiously, the SPC discourse legitimizes the extension of intellectual property principles to data. Policy documents like the *Opinions on Providing Judicial Services and Guarantees for Accelerating the Construction of a Unified National Market* highlight the tension between unifying market standards and preventing platform dominance (Supreme People's Court 2022c), showing how judicial activism remains integral to reconciling these competing demands.

The 2025 revision of the Anti-Unfair Competition Law shows the BCC theory's explanatory power (Standing Committee of the National People's Congress 2025). Article 13 prohibits platform operators from "using data, algorithms, technologies, or platform rules to obstruct or disrupt lawful network products or services" and from "improperly obtaining or using lawfully held data, infringing legitimate rights and disrupting competition." This illustrates how bidirectional conceptual coupling translates judicial reasoning – exemplified by the "conveyor model," which recognizes mutual legitimacy between incumbents and challengers – into statutory norms, framing data as both a legitimate right and a potential tool of unfair competition. The new Article 13 (3), banning the "abuse of platform rules," further reflects how judicial interpretations of "may" and "should" in CDA constrain platform-driven rule capture. This legislative evolution validates local judicial continuation discourse while reinforcing the SPC's judicial linkage discourse, ensuring coherence between DIPR judicial practice and national policy.

Overall, DIPR in judicial activism should be viewed as one "experimental field" in China's multilayered data governance. By shifting the focus from what data is to who defines and protects it, this study shows the judiciary's role in shaping DIPR as both a doctrinal innovation and a mechanism for institutional coordination.

**Acknowledgments:** The author acknowledges the valuable feedback provided by participants at the 2025 IPIRA Conference. The study also benefited from critical insights during institutional seminars at the University of Macau and Duke University. No Artificial Intelligence (AI) tools were used as credited authors; AI was only employed to assist with preliminary text revision under the author's direct oversight.
**Research ethics:** Not applicable.
**Author contributions:** The author has accepted responsibility for the entire content of this manuscript and approved its submission.
**Conflict of interest:** The author states no conflict of interest.
**Research funding:** None declared.
**Data availability:** Not applicable.

# Bionote

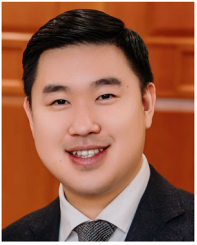

**Chanhou Lou**
Macau Fellow Researcher at the University of Macau, 2012 Office, E32, Avenida da Universidade, Taipa, 999078, Macau
**chlou@um.edu.mo**
**https://orcid.org/0009-0003-3494-1079**

Dr. Chanhou Lou is a Macau Fellow Researcher at the University of Macau, a Visiting Scholar at Cornell University, and a Digital Life Initiative (DLI) Fellow at Cornell Tech, specializing in IP law, data privacy law and legal history. He earned his Ph.D. in law *summa cum laude* from Tsinghua University, where he also completed his master's degree in International Intellectual Property. He is a legal scholar and data privacy specialist whose research focuses on the evolution of privacy law in contexts.